\documentclass{nature}
\usepackage{graphicx}
\makeatletter
\let\saved@includegraphics\includegraphics
\AtBeginDocument{\let\includegraphics\saved@includegraphics}
\renewenvironment*{figure}{\@float{figure}}{\end@float}
\makeatother
\usepackage{array,tabularx,booktabs} % in preamble
\newcolumntype{Y}{>{\centering\arraybackslash}X}
\usepackage{multirow}
\usepackage{subfigure}
\usepackage{mathtools}
\usepackage[colorlinks=true, linkcolor=blue,citecolor=black,urlcolor=blue]{hyperref}
\usepackage{authblk}
\usepackage[labelfont=bf]{caption}
\usepackage[figurename=Figure]{caption}
\usepackage{siunitx}
\usepackage[dvipsnames]{xcolor}
\usepackage{soul}
\usepackage{xfrac}
\usepackage{amsmath}
\usepackage{soul,xcolor}
\usepackage{lineno}
\usepackage{siunitx}
\usepackage{float}
\usepackage{hyphenat}

\graphicspath{{./figures/}}

\begin{document}

\setstcolor{red}

\title{\textbf{CNN-Based Classifier for Automated Identification of Magnetic States in Spin Dynamics Simulations}}
\author[1*]{Amal Aldarawsheh}
\author[2, 3*]{Ahmed Alia}
\author[1]{Stefan Blügel}

\affil[1]{Peter Gr\"{u}nberg Institute Forschungszentrum J\"{u}lich and JARA, D-52425 J\"{u}lich, Germany}
\affil[2]{Institute for Advanced Simulation (IAS-7), Forschungszentrum Jülich,  D-52425 J\"{u}lich, Germany}

\affil[3]{Faculty of Information Technology and Artificial Intelligence, An-Najah National University, Nablus, Palestine}

\affil[*]{a.aldarawsheh@fz-juelich.de, a.alia@fz-juelich.de}

\maketitle
%\linenumbers

\begin{abstract}

The identification and classification of different magnetic states  are essential for understanding the complex behavior of magnetic systems. Traditional approaches that rely on handcrafted features or manual inspection often fall short, particularly when dealing with subtle or topologically complex spin textures. In this study, we present an automated deep learning model that employs an EfficientNetV1B0  Convolutional Neural Network to classify nine distinct magnetic states, including both ferromagnetic (FM) and antiferromagnetic (AFM) spin textures such as AFM skyrmions  and AFM stripe domains. The spin configurations are generated through atomistic spin dynamics simulations using the \textit{Spirit} code, then visualized with VFRendering  to produce RGB images, which serve as inputs to the classification model.

To train and evaluate the model, we created a new dataset of manually labeled   RGB  images. Evaluation results show that the proposed model achieves an accuracy and F1-score of \SI{99}{\percent}, outperforming the other deep learning baselines evaluated in this study.
\end{abstract}

\section{Introduction}

The application of machine learning  and deep learning  in condensed matter physics has opened new pathways for understanding and engineering complex physical systems. These techniques are now widely used to represent quantum states~\cite{carleo2017solving}, discover phase transitions~\cite{rem2019identifying,wang2016discovering,van2017learning}, and classify magnetic phases across vast parameter spaces~\cite{broecker2017machine,ch2017machine,zhang2017machine,zhang2017quantum,carrasquilla2017machine,iakovlev2018supervised}. 
A domain where these methods have had significant impact is the study of topological magnetic textures, such as magnetic skyrmions. These nanoscale, swirling spin structures are stabilized by antisymmetric Dzyaloshinskii–Moriya interactions (DMI) and are topologically protected against small perturbations~\cite{muhlbauer2009skyrmion,romming2013writing,Fert2017magnetic,gobel2020beyond}. Their compact size, high stability, and efficient current-driven mobility make skyrmions highly promising for applications in spintronics, high-density memory, and neuromorphic computing~\cite{muhlbauer2009skyrmion,Sampaio2013,fert2013skyrmions,cortes2017thermal,moreau2016additive,wiesendanger2016nanoscale,zhang2015magnetic,puebla2020spintronic,song2020skyrmion,yokouchi2022pattern}. In addition to skyrmions, a rich variety of emergent magnetic states, including stripe domains, in-plane vortices, and N\'eel-type structures, have drawn attention for their technological relevance and for enriching our understanding of magnetic phase behavior in low-dimensional systems~\cite{Parkin2008,spethmann2021discovery,ma2018electric,grebenchuk2024topological,gobel2019magnetic,ezawa2011compact,capriotti1999long,white2007neel}. Classifying magnetic states from raw simulation data or experimental images remains a challenging task, particularly as simulations continue to generate increasingly complex spin textures and cover larger parameter spaces. This growing complexity underscores the need for automated classification methods capable of handling large-scale datasets and subtle variations in spin configurations.

Recently, Convolutional Neural Networks (CNNs) have demonstrated strong performance in image-based classification tasks. A key advantage of CNNs is their ability to automatically learn meaningful features from labeled images without the need for manual feature engineering~\cite{gu2018recent}. EfficientNet~\cite{tan2019efficientnet,tan2021efficientnetv2}, ResNet~\cite{he2016deep}, and MobileNet~\cite{howard2017mobilenets,sandler2018mobilenetv2} are popular families of CNN architectures. Among these, EfficientNetV1B0 stands out as a simple yet effective model, achieving high classification performance with  fewer parameters and lower computational cost~\cite{tan2019efficientnet, alia2024novel, alia2022hybrid}. These advancements in deep learning have motivated researchers to explore the use of CNNs for extracting physical insights from spin texture images generated through simulations or experiments.

Recent literature has demonstrated that treating spin configurations as image data enables effective training of CNNs, improving classification performance and  facilitating comparison with image-based experimental measurements~\cite{rem2019identifying, wang2020machine}. CNNs have been applied to classify complex spin textures like ferromagnetic (FM) spirals, bimerons, skyrmion crystals, and skyrmion gases with high accuracy~\cite{gomez2022cnn, araz2022identifying, feng2023classification, skyrnet2025 }, and have also been used to extract physical parameters such as chirality, magnetization, and Hamiltonian coefficients from spin configurations~\cite{wang2020machine, feng2023classification}. For example, Cimen et al.~\cite{skyrnet2025} designed a  customized Skyr-Net network  to classify seven  FM  micromagnetic spin textures from MuMax3 simulations, achieving near-perfect accuracy. Iakovlev et al.\cite{iakovlev2018supervised} employed basic machine-learning models to distinguish complex non-collinear magnetic states in 2D materials, reporting 91\% accuracy using the K-nearest neighbor algorithm. Similarly, Albarrán et al.\cite{gomez2022cnn} implemented a CNN to classify multiple spin phases, such as spirals, bimerons, skyrmion crystals, and paramagnetic states based on Monte Carlo simulations, reaching 97.5\% accuracy. Furthermore, Salcedo-Gallo et al.~\cite{salcedo2020deep} developed a CNN to classify FM, stripe, bimeron-skyrmion, and skyrmion phases from Monte Carlo simulations. Their  model achieved high accuracy and was used to construct phase diagrams by varying magnetic parameters. 

By contrast, machine-learning studies involving antiferromagnetism remain more limited in this image-based classification context and have mainly addressed related but distinct tasks, such as classification of magnetic order from electronic-structure descriptors, inference of magnetic structures from atomic coordinates, or prediction of the conditions for AFM skyrmion formation, rather than image-based multiclass classification of AFM spin textures themselves~\cite{jang2023classification,merker2022machine,saini2024machine}. Although some image-based studies have included an AFM ordered state as one class among several magnetic structures~\cite{hasib2021classification,bokul2020non}, to the best of our knowledge, previous image-based automatic classification studies have not explicitly treated AFM skyrmions and AFM stripe domains as separate target classes within a unified multiclass framework.

This gap is important because AFM textures are of considerable interest due to their faster spin dynamics, absence of stray fields, and potential for high-density, low-power spintronic devices~\cite{gomonay2016high,Jungwirth2016,baltz2018antiferromagnetic,aldarawsheh2022emergence,aldarawsheh2023spin}. Nevertheless, most existing studies focus mainly on FM systems and a narrow set of phases, typically uniform FM states, spirals, and skyrmions.

Moreover, a key limitation of many previous efforts lies in their use of low-dimensional or simplified representations of spin textures, such as scalar observables or $z$-projected components~\cite{iakovlev2018supervised,salcedo2020deep,feng2023classification}, which obscure critical features encoded in the full three-dimensional spin structure. These limitations underscore the need for models that are trained on physically consistent, atomistic data capable of capturing the full diversity of magnetic textures.

\begin{figure}[h!] 
    \centering
 \includegraphics[width=1\textwidth]{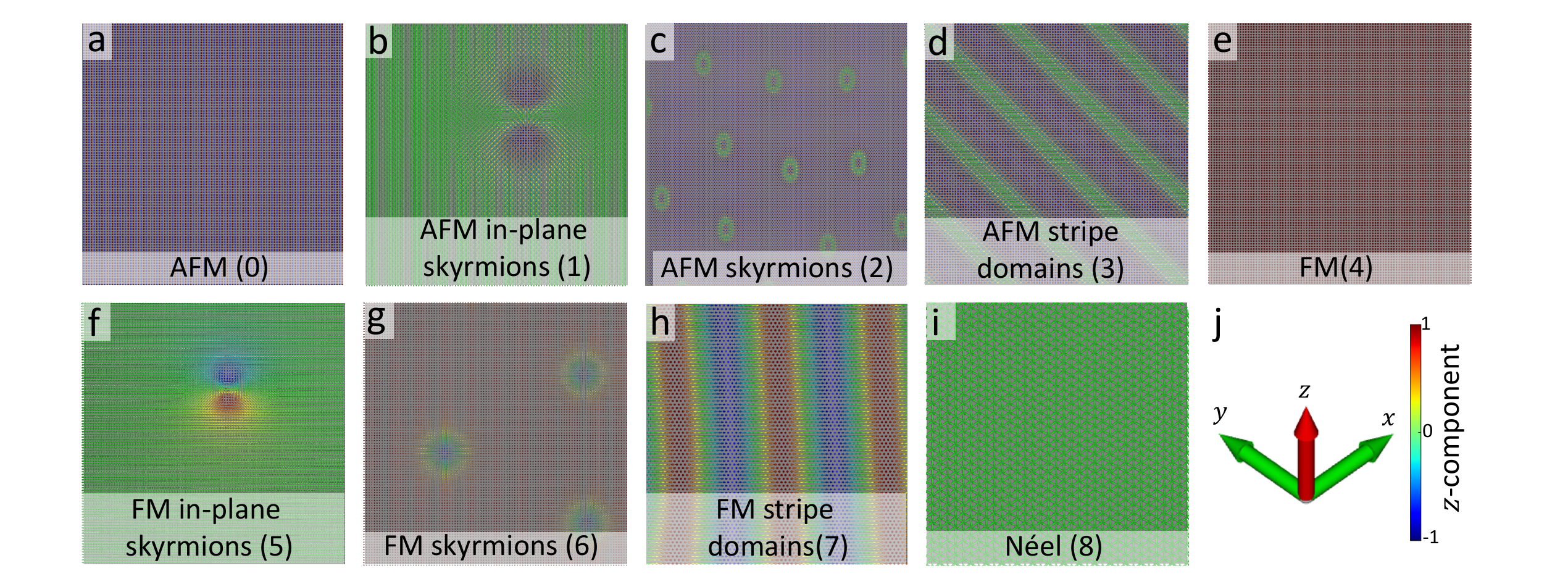}
    \caption{\textbf{Sample snapshots from the dataset.}  \textbf{a} AFM (0), \textbf{b} AFM in-plane skyrmions (1), \textbf{c} AFM skyrmions (2), \textbf{d} AFM stripe domains (3), \textbf{e} FM (4), \textbf{f} FM in-plane skyrmions (5), \textbf{g} FM skyrmions (6), \textbf{h} FM stripe domains (7),  and \textbf{i} N\'eel state (8), \textbf{j} visualization of the color code used.}
    \label{fig:states}
\end{figure}

In this study, we present an automated CNN-based model for classifying visualized spin-configuration files into nine distinct magnetic states. These configurations are generated through atomistic spin dynamics simulations and subsequently visualized as RGB images using the VFRendering tool~\cite{VFRendering2023}. Figure~\ref{fig:states} illustrates one representative example for each magnetic state. For the classification task, we adapt and train the EfficientNetV1B0 architecture. The original network is integrated with a fully connected layer followed by a Softmax activation function with nine output neurons, forming the final classification layer. The adapted architecture is then trained from scratch on a newly labeled image dataset, where each image represents a spin configuration.

Compared with many earlier CNN-based studies~\cite{iakovlev2018supervised,salcedo2020deep,gomez2022cnn,araz2022identifying,feng2024classification,skyrnet2025}, our proposed model incorporates both FM and AFM spin textures, including in-plane and out-of-plane configurations. In particular, it explicitly includes AFM skyrmions and AFM stripe domains as dedicated  classes within a unified multiclass CNN framework trained on atomistic spin-texture images. In addition, the present study considers nine magnetic states across different lattice types using atomistic spin-dynamics data, thereby broadening prior CNN-based magnetic-state classification studies to a nine-class setting.\\

\section{Model and methods}
\label{methods}

In this section, we first present the architecture of the classification model, which is designed to categorize input RGB images into nine magnetic states (see Fig.~\ref{fig:states}). This is followed by a description of the preparation of the labeled image dataset required for training and evaluating the model. Further details are provided in the following subsections.

\subsection{Adapted EfficientNetV1B0-based classification model.}
CNNs are a class of deep learning  inspired by the organization of the animal visual cortex and are widely used in vision-related applications, including image classification~\cite{younesi2024comprehensive}. One of their main advantages is the ability to automatically and adaptively learn relevant features from data without requiring handcrafted feature engineering~\cite{gu2018recent}. A typical CNN architecture consists of repeated convolution and pooling layers for feature extraction, followed by one or more  fully connected layers that map the extracted features to output neurons corresponding to classification categories~\cite{yamashita2018convolutional}.

The design of the CNN architecture plays a crucial role in improving model performance. Several efficient and high-performing architectures have been developed, such as  InceptionResNetV2~\cite{szegedy2017inception}, DenseNet121~\cite{huang2017densely}, MobileNet~\cite{howard2017mobilenets}, MobileNetV2~\cite{sandler2018mobilenetv2}, 
MobileNetV3Small~\cite{howard2019searching}, 
ResNet50~\cite{he2016deep}, ResNet101~\cite{he2016deep},  Xception~\cite{chollet2017xception}, and  EfficientNetV1B0~\cite{tan2019efficientnet}. 
This final architecture stands out for its simplicity and efficiency, delivering high performance on challenging benchmark image classification tasks~\cite{alia2022hybrid,alia2024novel}. It achieves these results with significantly fewer parameters than comparable models and reduced training costs.
Additionally, the experiments presented in this article (see Section~\ref{sec:experimentsandresults}) demonstrate that EfficientNetV1B0 achieves the highest accuracy compared to other popular CNN architectures in the task of magnetic state classification. As a result, our model is built based on the EfficientNetV1B0 CNN architecture.

\begin{figure}[h!] 
    \centering
    \includegraphics[width=0.9\textwidth]{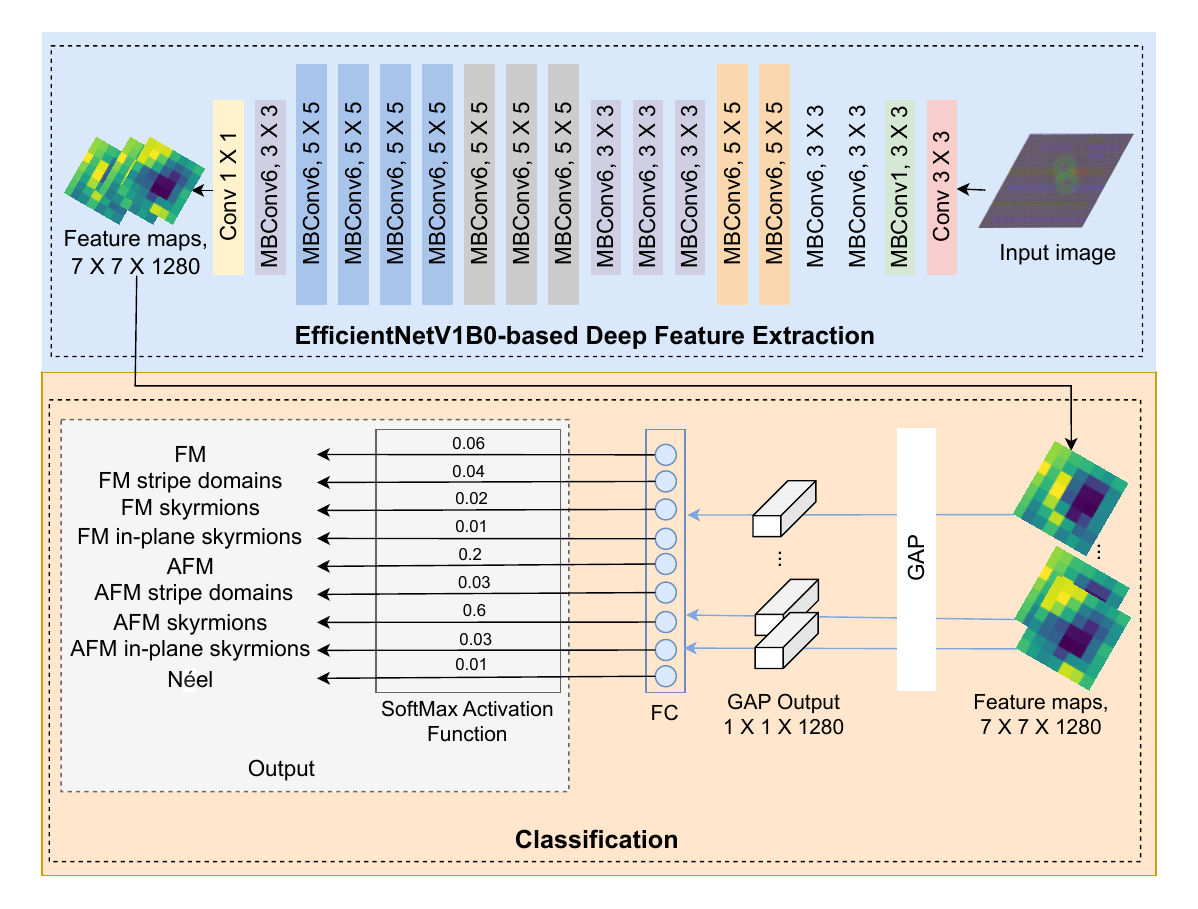}
    \caption{The architecture of the proposed model.}
    \label{fig:cnn-architecture2}
\end{figure}

\begin{figure}[h!] 
    \centering
    \includegraphics[width=0.5\textwidth]{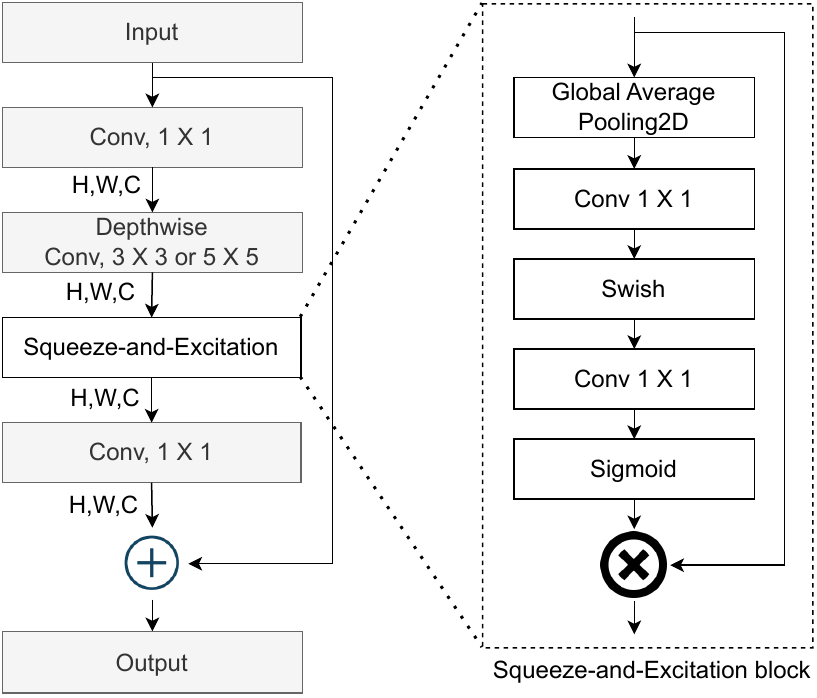}
    \caption{The architecture of MBConv block.}
    \label{fig:mbconv}
\end{figure}

As illustrated in Fig.~\ref{fig:cnn-architecture2},  the feature extraction component of EfficientNetV1B0 is used to capture informative representations from each magnetic state image, while the classification part is adapted to support nine-class classification. 
For feature extraction (Fig.~\ref{fig:cnn-architecture2}, top box), EfficientNetV1B0 begins with a standard \(3 \times 3\) convolutional operation that processes the input magnetic state image of size \(224 \times 224 \times 3\). This is followed by 16 Mobile Inverted Bottleneck Convolution (MBConv) blocks to extract the deep features.  MBConv blocks form the core of the architecture, and include one MBConv1 block with a \(3 \times 3\) kernel, six MBConv6 blocks with \(3 \times 3\) kernels, and nine MBConv6 blocks with \(5 \times 5\) kernels. Figure~\ref{fig:mbconv} shows the architecture of MBConv. Each MBConv block consists of several operations: it starts with a pointwise \(1 \times 1\) convolution that expands the number of channels, followed by a depthwise convolution (\(3 \times 3\) or \(5 \times 5\)) that applies a spatial filter to each channel independently, thereby reducing the number of learnable parameters. Batch normalization and the Swish activation function~\cite{ramachandran2017searching} are applied after each convolution to stabilize training and promote non-linearity.

A key innovation in MBConv blocks is the integration of the Squeeze-and-Excitation  module~\cite{hu2018squeeze}, which enhances the network’s representational capacity by learning dynamic channel-wise weights. This module first applies global average pooling to compress spatial information into a channel descriptor, followed by two fully connected layers using Swish and Sigmoid activations to compute attention weights for each channel. These learned weights are multiplied by the original feature maps to emphasize the most informative features. Finally, a second \(1 \times 1\) pointwise convolution reduces the feature depth back to the original dimension, and a residual connection is added to preserve gradient flow and mitigate vanishing gradient issues during training. The main difference between MBConv6 and MBConv1 is the depth of the block and the number of operations performed in each block; MBConv6 is six times that of MBConv1. Note that MBConv6, $ 5 \times  5$ performs the identical operations as MBConv6, $ 3 \times  3$, but MBConv6, $ 5 \times  5$ applies a kernel size of $ 5 \times  5$, while MBConv6, $ 3 \times  3$  uses a kernel size of $ 3 \times  3$.

This model supports nine-class classification by employing a $1 \times 1$ convolutional layer, followed by 2D global average pooling, a fully connected layer with nine output neurons, and a Softmax activation function (Fig.~\ref{fig:cnn-architecture2}, bottom box). To enable the model to classify magnetic states into nine distinct classes, the adapted EfficientNetV1B0 CNN architecture must be trained using a labeled dataset consisting of RGB images corresponding to the nine magnetic state categories. The next section describes the preparation of this required dataset.

\subsection{Dataset preparation.}
\label{dataset}
\begin{figure}[h!] 
    \centering
 \includegraphics[width=\textwidth]{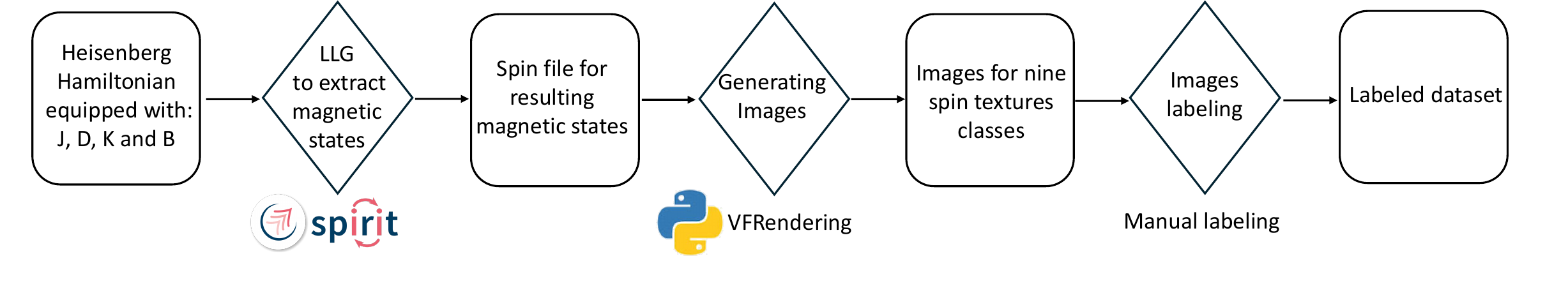}
    \caption{\textbf{Dataset preparation diagram.}}
    \label{fig:data-set-preparation}
\end{figure}

Figure~\ref{fig:data-set-preparation} shows the main steps for creating the labeled dataset used to train and evaluate the adapted EfficientNetV1B0 model, as well as the other models included in the evaluation. To generate the magnetic states that represent the samples in this dataset, we employed a classical atomistic spin model defined by the two-dimensional Heisenberg Hamiltonian with the inclusion of Heisenberg exchange interactions ($J$), Dzyaloshinskii–Moriya interaction (DMI), uniaxial anisotropy $K$, and Zeeman coupling:

\begin{equation}
 \mathcal{H}= -\sum\limits_{\langle ij \rangle} J_{ij}\;\boldsymbol{n}_{i}\cdot \boldsymbol{n}_{j}  
 - \sum\limits_{\langle ij \rangle}\mathbf{D}_{ij}\cdot (\boldsymbol{n}_{i}\times \boldsymbol{n}_{j})
 - K\sum\limits_{i}  (n_i ^z)^2  
 - \sum\limits_{i}  \mathbf{B} \cdot \mathbf{n}_i,
\end{equation}
where $i$ and $j$ are site indices carrying each magnetic moment.  $\textbf{n}$ is a unit vector of the magnetic moment. $J_{ij}$ is the Heisenberg exchange coupling strength, being $<$ 0 for AFM interaction, between an atom on site $i$ and another atom on site $j$. A similar notation is adopted for the DMI vector $\textbf{D}$ and the magnetic anisotropy energy $K$. The latter favors the out-of-plane orientation of the magnetization if being $>$ 0, or in-plane orientation if being $<$ 0. Finally \textbf{B} is the external magnetic field.
 The Heisenberg exchange interaction values  were adjusted to model both FM and AFM systems. In the FM case, we included the first nearest neighbors with FM coupling, while in the AFM case, we included up to third-nearest-neighbor interactions, particularly on triangular lattice, following our earlier findings on the emergence of intrinsic AFM skyrmions~\cite{aldarawsheh2022emergence,aldarawsheh2023spin}. Moreover, different DMI orientations were employed to stabilize various types of spin textures, including N\'eel-type, Bloch-type, and antiskyrmion configurations. The sign of the anisotropy term $K$ was varied to support both in-plane and out-of-plane spin alignments. Additionally, a magnetic field along the $z$-direction was applied in selected FM simulations to enhance the stability of skyrmions.
To avoid the nonuniversal effects of boundary conditions, we used a supercell on the 200 $\times$ 200 spin lattice containing 40,000 sites with periodic boundary conditions. Simulations were performed using the \textit{Spirit} atomistic spin dynamics package~\cite{muller2019spirit}, which integrates the Landau–Lifshitz–Gilbert (LLG) equation to evolve spin configurations toward their ground states. We considered four distinct lattice geometries; square, triangular, rhombic,  and rectangular, to ensure broad coverage of structural symmetries and magnetic interaction scenarios.    The Hamiltonian parameters were systematically varied over broad intervals to stabilize the nine targeted magnetic states across the considered lattice geometries (Fig.~\ref{fig:states}). Representative parameter ranges for each magnetic state and lattice type are provided in Supplementary Table~S1.

  Following convergence, the underlying spin configurations  generated on 
200 $\times$ 200 lattices were  visualized using the VFRendering script~\cite{VFRendering2023}, which converts each spin-configuration file  into an RGB image. These images were then resized to  224 $\times$ 224 pixels to match the CNN input size. In our RGB images, the arrow color encodes the out-of-plane ($z$) component of the spin, ranging from red (up) to blue (down), as shown in Fig.~\ref{fig:states}j, while the arrow orientation reflects the in-plane (x,y) direction, thereby preserving the full 3D spin vector information. This full-vector representation is particularly important for the in-plane cases of FM, AFM, and N\'eel states, for which the $z$-projection alone   provides only limited visual contrast, whereas the arrow orientation  preserves the in-plane spin direction pattern needed to distinguish these states. Supplementary Fig.~S1 and Supplementary Table~S2 compare the full-arrow and $z$-projection only representations and provide the corresponding quantitative similarity analysis.

To ensure consistency across the dataset and to avoid uncontrolled visual variation in the CNN inputs, the rendering parameters were kept fixed throughout image generation, including the colormap, arrow representation, and background settings. 

 After image generation, the images  were manually labeled into nine distinct magnetic states mentioned earlier: AFM ground state (0), AFM in-plane skyrmions (1), AFM skyrmions (2), AFM stripe domains (3),  FM ground state (4),  FM in-plane skyrmions (5), FM skyrmions (6), FM stripe domains (7), and N\'eel state (8).  The resulting dataset consists of 6,503 RGB images distributed into nine classes representing distinct magnetic states.

To create holdout data for model training and evaluation,  the full labeled dataset was randomly divided class-wise  into three subsets: 70\% for training, 15\% for validation, and 15\% for testing. This split ratio is widely adopted in the deep learning community and provides a balanced approach for training and evaluating model performance~\cite{genc2019optimal}. Table~\ref{tab:data-set-distr} presents the number of samples in each magnetic state class across the training,  validation, and test sets.

\begin{table} 
\centering
\footnotesize
\caption{Distribution of spin-texture images across the nine magnetic state classes for training,  validation, and test sets.}
\label{tab:data-set-distr}
\begin{tabular}{lccc c}
\hline
\textbf{Class} & \textbf{Training} & \textbf{Validation} & \textbf{Test} & \textbf{Total} \\
\hline
AFM                  & 500 & 115 & 105 & 720 \\
AFM in-plane skyrmions  & 627 & 146 & 128 & 901 \\
AFM skyrmions           & 627 & 145 & 128 & 900 \\
AFM stripe domains             & 482 & 110 & 100 & 692 \\
FM                   & 525 & 115 & 110 & 750 \\
FM in-plane skyrmions   & 522 & 121 & 105 & 748 \\
FM skyrmions            & 520 & 121 & 105 & 746 \\
FM stripe domains    & 523 & 117 & 110 & 750 \\
N\'eel                    & 206 &  46 &  44 & 296 \\
\hline
\textbf{Total}           & 4532 & 1036 & 935 & 6503 \\
\hline
\end{tabular}
\end{table}

\subsection{Evaluation metrics.}

To evaluate the performance of our proposed model, we employed macro-accuracy and macro F1-score metrics that are well-suited for multi-class problems. The following provides a detailed explanation of these metrics:

\begin{enumerate}

    \item \textbf{Macro-accuracy,}  calculated by computing the accuracy independently for each class and then averaging the results. This ensures equal contribution from all classes, regardless of class imbalance. It is particularly appropriate in our case since some magnetic states are underrepresented in the dataset (e.g., N\'eel state), and overall accuracy alone may mask poor performance on minority classes.

    \item \textbf{F1-score} is the harmonic mean of precision and recall:
    \[
    \text{F1-score} = 2 \cdot \frac{\text{Precision} \cdot \text{Recall}}{\text{Precision + Recall}}.
    \]
    In other words, F1-score provides a balanced measure of classification performance by incorporating precision and recall.  

    Precision is defined as the proportion of correctly predicted instances of a class among all instances predicted as that class:
    \[
    \text{Precision} = \frac{\text{True Positives}}{\text{True Positives + False Positives}}, 
    \]
    and Recall (or sensitivity) is defined as the proportion of correctly predicted instances of a class among all actual instances of that class:
    \[
    \text{Recall} = \frac{\text{True Positives}}{\text{True Positives + False Negatives}}.
    \]

    In our analysis, we use the \textbf{macro F1-score}, which averages the F1-scores of all classes equally, providing a comprehensive view of the model's performance across the nine magnetic states.
\end{enumerate}

In summary, these metrics were chosen to ensure that our model performs consistently across all classes, and not just those that dominate the dataset.

\begin{table}
\centering
\caption{The hyperparameter values used in the training process}
\label{tab:hyperparameters}
\begin{tabular}{ll}
\hline
\textbf{Parameter}     & \textbf{Value}            \\
\hline
Optimizer              & Adam                      \\
Loss function          & Categorical cross-entropy \\
Learning rate          & 0.001                     \\
Batch size             & 32                        \\
Epochs                 & 50                        \\
\hline
\end{tabular}
\end{table}

\section{Experiments and results}
\label{sec:experimentsandresults}

This section presents the implementation details and training parameter settings. It also discusses the results of various experiments conducted to evaluate the performance of the proposed classifier.

\subsection{Implementation details and hyperparameters.}

All experiments and implementations in this study were conducted on Google Colaboratory Pro, utilizing a 15 GB NVIDIA GPU and 12.7 GB of system RAM, utilizing Python along with Keras and TensorFlow libraries. Table~\ref{tab:hyperparameters} shows hyperparameters used during the training process. The default values commonly used for CNN training on ImageNet in Keras were adopted for the learning rate and batch size. Additional parameters were fine-tuned through experimentation to achieve optimal performance with magnetic states classification. The model was trained for 50 epochs with early stopping after 10 epochs being enabled to avoid overfitting.

\subsection{Experimental results.}

Here, a series of experiments  were conducted to evaluate the performance of the proposed classification model. The model was trained on the training and validation sets, and its performance was assessed on the test set using the macro accuracy, recall, precision and macro F1-score metrics. For comparative analysis, the proposed EfficientNetV1B0-based model was compared with eight widely used CNN architectures: InceptionResNetV2~\cite{szegedy2017inception}, DenseNet121~\cite{huang2017densely}, MobileNet~\cite{howard2017mobilenets}, MobileNetV2~\cite{sandler2018mobilenetv2}, MobileNetV3Small~\cite{howard2019searching}, ResNet50~\cite{he2016deep}, ResNet101~\cite{he2016deep}, and Xception~\cite{chollet2017xception}. Each architecture was adapted to the nine-class classification task using the same classification head. To ensure a fair comparison, all models were implemented using the same hardware and software environments,  hyperparameters  (Table~\ref{tab:hyperparameters}), evaluation metrics, and dataset (Table~\ref{tab:data-set-distr}).

\begin{table}
\centering
\small
\setlength{\tabcolsep}{4pt}
\renewcommand{\arraystretch}{0.95}
\caption{Comparison of the proposed EfficientNetV1B0-based model with eight widely used CNN architectures. Reported metrics include the total number of trainable parameters, macro accuracy, precision, recall, and F1-score for each model.}
\label{tab:cnn-performance-comparison}
\begin{tabular}{clccccc}
\hline
  & \textbf{Model} & \textbf{Trainable Parameters} & \textbf{Macro Accuracy} & \textbf{Precision} & \textbf{Recall} & \textbf{F1-Score} \\
\hline
  & EfficientNetV1B0   & $\sim$ 4.02 M  & 99\% & 99\% & 99\% & 99\% \\
  & MobileNet          & $\sim$ 3.22 M  & 97\% & 97\% & 97\% & 97\% \\
  & ResNet50           & $\sim$ 23.55 M & 94\% & 93\% & 94\% & 94\% \\
  & Xception           & $\sim$ 20.83 M & 97\% & 97\% & 97\% & 97\% \\
  & MobileNetV2        & $\sim$ 2.24 M  & 66\% & 78\% & 65\% & 65\% \\
  & DenseNet121        & $\sim$ 6.96 M  & 97\% & 98\% & 97\% & 97\% \\
  & ResNet101          & $\sim$ 42.57 M & 96\% & 96\% & 96\% & 96\% \\
  & InceptionResNetV2  & $\sim$ 54.29 M & 97\% & 97\% & 97\% & 97\% \\
  & MobileNetV3Small   & $\sim$ 0.93 M  & 97\% & 97\% & 97\% & 97\% \\
\hline
\end{tabular}
\end{table}

\begin{figure}[h!] 
    \centering
    \includegraphics[width=0.5\textwidth]{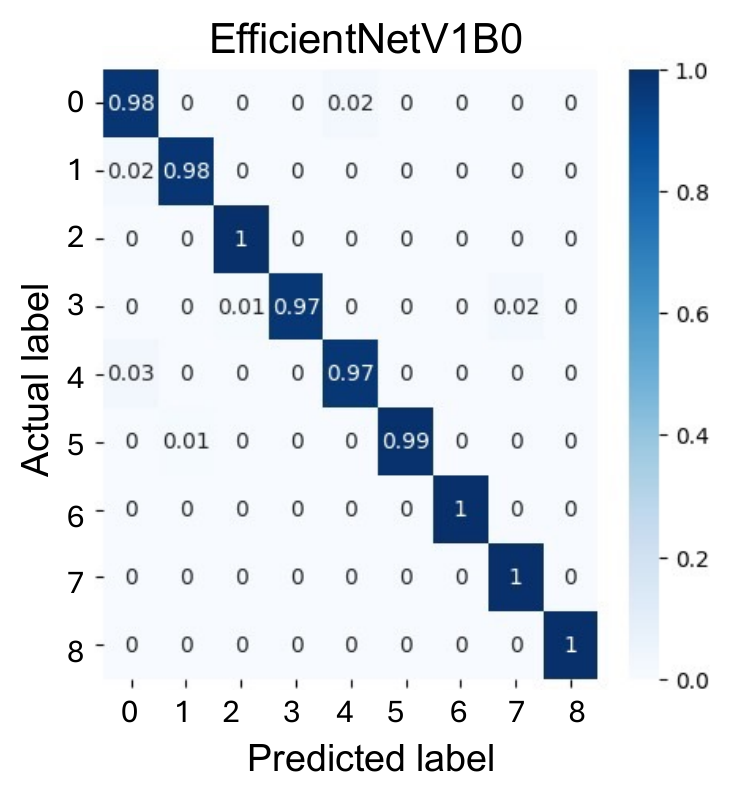}
    \caption{Confusion matrix of the EfficientNetV1B0-based classification model. The class labels 0–8 correspond to AFM, AFM in-plane skyrmions, AFM skyrmions, AFM stripe domains, FM, FM in-plane skyrmions, FM skyrmions, FM stripe domains, and the N\'eel state, respectively.}
    \label{fig:conf-effnet}
\end{figure}

As shown in Table~\ref{tab:cnn-performance-comparison} and in Fig.~\ref{fig:conf-effnet}, the EfficientNetV1B0-based model  performed well in classifying the nine magnetic states in our dataset. It achieved a macro accuracy, precision, recall, and F1-score of 99\%, indicating high classification performance on the held-out test set spanning the nine magnetic states considered in this study. 
 As illustrated by the confusion matrix in Fig.~\ref{fig:conf-effnet}, the model correctly predicted nearly all samples across classes, including challenging categories such as  AFM and FM skyrmions, as well as their in-plane variants. Misclassifications were rare and primarily occurred between states with subtle differences, such as in-plane AFM and FM  ground states or stripe domains.
 
 To further assess the robustness of the reported performance, we considered the effect of class imbalance in the dataset. The class distribution is not uniform, with the N\'eel class representing the smallest category (Table~\ref{tab:data-set-distr}). We therefore examined the confusion matrix in Fig.~\ref{fig:conf-effnet} together with the class-wise recall values and their 95\% Wilson confidence intervals~\cite{brown2001interval} shown in Table~\ref{tab:ci}. The EfficientNetV1B0 model showed consistently high performance across all classes, with class-wise recall ranging from 97\% to 100\%. In particular, the N\'eel class achieved 100\% recall on the test set (44/44), corresponding to a 95\% Wilson confidence interval of 92--100\%. Thus, within the present test set, no reduction in class-wise recall was observed for the minority class.

\begin{table} 
\centering
\footnotesize
\caption{ Class-wise recall (\%) of the EfficientNetV1B0 model on the test set, with 95\% Wilson confidence intervals.}
\label{tab:ci}
\begin{tabular}{lccc c}
\hline
\textbf{Class} & \textbf{Samples} & \textbf{Recall (\%)} & \textbf{95\% Wilson Confidence Interval (\%)}  \\
\hline
AFM                & 105  &98  & 93.3 -- 99.5  \\
AFM in-plane skyrmions  & 128 & 98 & 93.3 -- 99.2  \\
AFM skyrmions         &128  & 100 & 97.1 -- 100  \\
AFM stripe domains & 100             & 97 & 91.5 -- 99 \\
FM  & 110                 &97 &  92.3 -- 99.1  \\
FM in-plane skyrmions & 105   & 99 & 94.8 -- 99.8  \\
FM skyrmions  & 105           & 100 & 96.5 -- 100  \\
FM stripe domains & 110    & 100 & 96.6 -- 100  \\
N\'eel & 44                   & 100 &  92 -- 100  \\
\hline
  \\
 
\end{tabular}
\end{table}
 
\begin{figure}[h!]
    \centering
    \includegraphics[width=\textwidth]{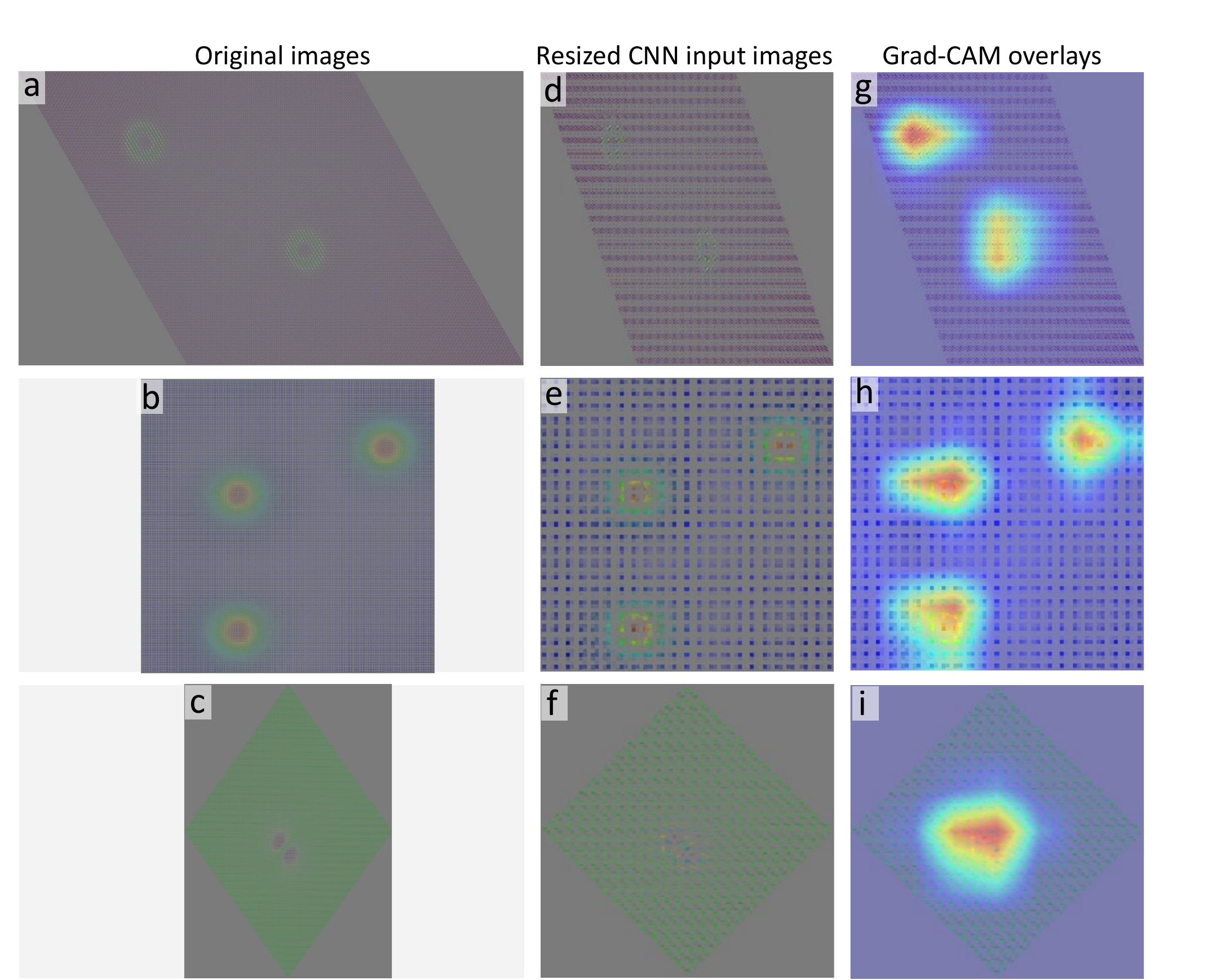}
    \caption{Qualitative Grad-CAM visualizations for three representative samples obtained with the EfficientNetV1B0-based CNN classifier. The first column (a–c) shows the original rendered images displayed with preserved aspect ratios, the second column (d–f) shows the resized images used as CNN input (224 $\times$ 224 pixels), and the third column (g–i) presents the corresponding Grad-CAM overlays. Brighter regions indicate stronger model attention and are concentrated on signal-bearing regions of the spin configurations.}
    \label{fig:grad-cam}
\end{figure}

Additionally, Figure~\ref{fig:grad-cam}  provides qualitative insight into the features   learned by the EfficientNetV1B0\hyp{}based CNN classifier. For the representative samples, the Grad-CAM~\cite{selvaraju2017grad} responses are concentrated on signal-bearing regions of the spin configurations and are not dominated by empty background areas or image boundaries. Since the rendered dataset was generated using fixed visualization parameters, this qualitative analysis suggests that the classifier is primarily responding to task-relevant spatial features rather than to obvious artifacts of the image-generation pipeline.\\

\begin{figure}[h!] 
    \centering
\includegraphics[width=\textwidth]{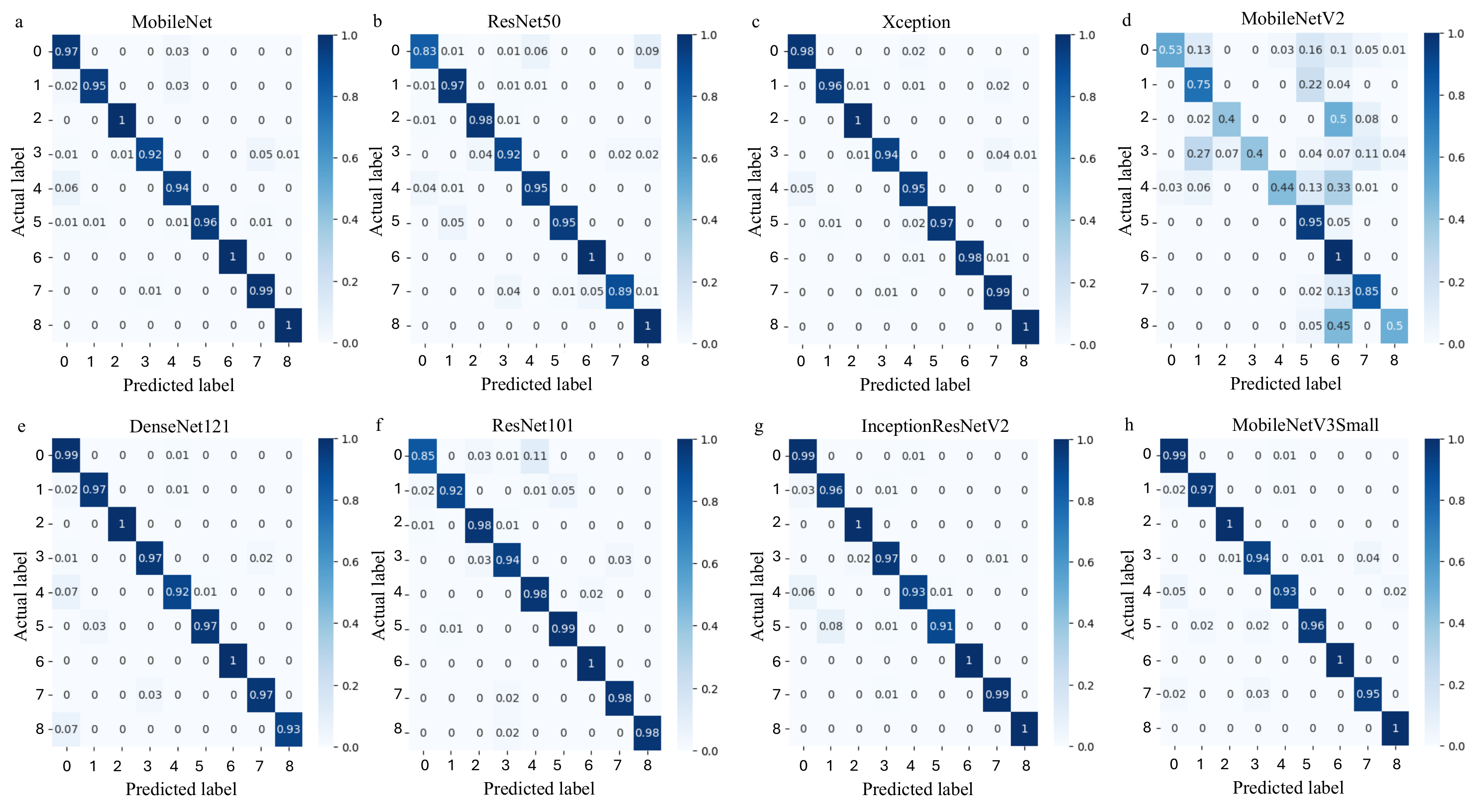}
   \caption{Confusion matrices of the other  CNN models trained on the magnetic state classification dataset. Each matrix illustrates the model's classification performance across the nine target classes: AFM (0), AFM in-plane skyrmions(1), AFM skyrmions(2), AFM stripe domains(3), FM(4), FM in-plane skyrmions(5), FM skyrmions(6), FM stripe domains(7), and the N\'eel state(8). Panels correspond to \textbf{a} MobileNet, \textbf{b} ResNet50,  \textbf{c} Xception, \textbf{d} MobileNetV2, \textbf{e} DenseNet121, \textbf{f} ResNet101, \textbf{g} InceptionResNetV2, and \textbf{h} MobileNetV3Small. Diagonal values indicate correct predictions, while off-diagonal values indicate misclassifications.}
    \label{fig:confusion-matrix}
\end{figure}

The comparison with other CNN architectures further shows that EfficientNetV1B0 achieved the best overall combination of accuracy, precision, recall, F1-score, and trainable-parameter count among the evaluated models, indicating that it is effective for distinguishing the visual patterns present in the rendered magnetic-texture dataset used in this study.
These results are consistent with the ability of EfficientNetV1B0 to extract spatial features efficiently from spin-texture images.
While several models, such as DenseNet121, MobileNet, and InceptionResNetV2, also showed comparable performance, others like MobileNetV2 lagged  behind, particularly in recall and F1-score. Figure~\ref{fig:confusion-matrix} presents the confusion matrices for all models, providing a detailed breakdown of their classification strengths and weaknesses across the nine magnetic state classes.

Because the workflow starts from spin-configuration  files before visualization, it is not inherently restricted to \textit{Spirit}-generated inputs and may be extended to outputs from other simulation tools that provide OVF spin files. As a preliminary external-code check, we evaluated the trained model on 20 rendered images generated from MuMax3~\cite{vansteenkiste2014design} spin-configuration files, comprising 10 FM skyrmion states, 5 FM stripe-domain states, and 5 FM states, all of which were classified correctly. This provides an initial indication of cross-code compatibility for these classes. Overall, these results support the use of CNN-based methods to automate the analysis of rendered spin-dynamics configurations, enabling more scalable, efficient, and objective interpretation of magnetic phenomena.

\section{Conclusion, limitations and future work}

In this work, we introduced an automated deep learning framework for classifying nine distinct magnetic states, including both FM and AFM spin textures, generated from atomistic spin dynamics simulations. The framework is based on an adapted EfficientNetV1B0 CNN architecture with a nine-class output layer and was trained from scratch on a newly generated labeled dataset of spin-texture images. Evaluated on data spanning multiple lattice geometries and magnetic regimes, the framework explicitly incorporates AFM skyrmions and AFM stripe domains within a unified classification model. Under the controlled conditions considered here, the model accurately classified rendered images of simulated in-plane and out-of-plane magnetic configurations, achieving \SI{99}{\percent} accuracy and macro F1-score.

A limitation of the present study is that it relies on synthetic, noise-free images generated from simulated spin configurations using a fixed rendering protocol. Accordingly, the reported results do not establish robustness to  variations in visualization settings such as spin projection, colormap, arrow representation, or background fraction. Because the CNN learns from spatial and visual features in the rendered images, such variations may alter the image statistics and affect classification performance. In addition, many experimental techniques provide only partial magnetic information, for example predominantly in-plane or out-of-plane contrast depending on the imaging modality. Extending the present framework to experimental data will therefore require modality-specific datasets together with retraining or fine-tuning of the classifier. Furthermore, the present study considers ground-state and metastable configurations and does not address finite-temperature effects, where thermal fluctuations may lead to less well-defined spin textures and potentially affect classification performance.

Future work will therefore focus on extending the framework to experimental image data through modality-specific dataset construction and retraining, as well as to projection-limited inputs in which only partial magnetic information is available. It will also be important to evaluate robustness with respect to visualization parameters such as colormap and background fraction, and to further integrate explainable AI techniques to better interpret the model’s classification decisions.  Overall, the present study demonstrates that a unified CNN-based framework can accurately classify a diverse set of simulated spin textures, including both FM and AFM configurations.

\section*{Acknowledgments}
The authors thank Nikolai Kiselev and Thorben Pürling for fruitful discussions. This work was financially supported by the Deutsche Forschungsgemeinschaft (DFG) through CRC 1238 (Project C01) and the European Research Council grant 856538 (project "3D MAGIC").

\section*{Author Contributions}
A.A. conceived the idea and designed the study. She generated and curated the dataset, performed data labeling, and implemented, modified, and trained the convolutional neural network models. A.Alia contributed to model development, implementation, and performance evaluation. S.B. supervised the project and contributed to the conceptual development of the work. All authors discussed the results and contributed to the writing and revision of the manuscript.

\section*{Competing Interests}
The authors  declare no competing interests.

\section*{Data Availability}
The datasets generated and analyzed during the current study, together with the trained models, are available at: 
\url{https://github.com/Amalaldarawsheh/CNN-MagSpin}
\section*{References}
\bibliographystyle{naturemag}
\bibliography{references}

\end{document}